\documentstyle[aps,manuscript]{revtex}
\begin{document}
\title{Three-dimensional harmonic oscillator and time evolution in quantum mechanics}
\draft
\author{Pavel~Kundr\'{a}t
\footnote{E-mail: Pavel.Kundrat@fzu.cz},
Milo\v{s}~Lokaj\'{\i}\v{c}ek \footnote{E-mail: lokaj@fzu.cz}}
\address{Institute~of~Physics, Academy~of~Sciences of~the~Czech~Republic,
Na~Slovance~2, CZ-182~21~Praha~8, Czech~Republic} \maketitle

\begin{abstract}
The problem of defining time (or phase) operator for
three-dimensional harmonic oscillator has been analyzed. A new
formula for this operator has been derived. The results have been
used to demonstrate a possibility of representing
quantum-mechanical time evolution in the framework of an extended
Hilbert space structure. Physical interpretation of the extended
structure has been discussed shortly, too.
\\

\end{abstract}

\section{Introduction}
\label{sec1}

The description of time evolution in the framework of the standard
quantum mechanics represents a problem that has not yet been
satisfactorily solved, even if the given theory is based in
principle on the time-dependent Schr\"{o}dinger equation. The
first attempt of defining an operator, the expectation values of
which corresponded to increasing time, was undertaken by Dirac
\cite{Dirac} already in 1927. He tried to express the operator
fulfilling the condition
\begin{equation}
  i [H,T] = 1                      \label{tim}
\end{equation}
for linear harmonic oscillator (or electromagnetic field) with the
help of annihilation and creation operators. He defined the
operator
\begin{equation}
  {\mathcal{E}} \;=\;  \exp(-i\omega T)         \label{epx}
\end{equation}
fulfilling corresponding commutation relations with Hamiltonian
\begin{equation}  \label{[H,E]}
  [H,{\mathcal{E}}]=-\omega{\mathcal{E}}, \qquad
           [H,{\mathcal{E}}^\dagger]=\omega {\mathcal{E}}^\dagger.
\end{equation}

However, it was shown later by Susskind and Glogover \cite{SG}
that the required unitarity of this operator did not hold for all
states of the standard Hilbert space, as
${\mathcal{E}}^\dagger{\mathcal{E}} |0 \rangle = 0$; i.e., the
condition has not been fulfilled for the state of minimum energy
(vacuum state). Many authors have tried to remove this difficulty
and to find a suitable modification of the standard Hilbert space
enabling to define a unitary phase exponential operator
${\mathcal{E}}$; see, e.g., the review of Lynch \cite{Lynch}.

The situation has been complicated due to the other argument
presented by Pauli \cite{Pauli} already in 1933. He showed that
the existence of time operator $T$ fulfilling Eq. (\ref{tim}) in
the standard Hilbert space (i.e., in the Hilbert space spanned on
a simple basis of Hamiltonian eigenvectors) required for
Hamiltonian $H$ to possess continuous energy spectrum from the
whole interval $E\in (-\infty,+\infty)$, which contradicts the
energy non-negativity (or at least energy limited from below).

It has been generally assumed that both these problems have
followed from one common source and may be solved with the help of
one approach. However, in fact they represent two different
problems, which may manifest better in the three-dimensional case
than in the linear one.

In Sec. II we will repeat shortly the problem concerning the case
of linear harmonic oscillator; some approaches trying to solve the
question of non-unitarity of exponential phase operator will be
mentioned, too. The time (phase) operator for a three-dimensional
harmonic oscillator will be derived newly in Sec. III. In Sec. IV
the doubled Hilbert space will be constructed for this system,
using the approach of Fain \cite{Fain} and Newton \cite{Newton}.
This approach has been originally proposed for removing the
problem in the case of linear harmonic oscillator; however, the
physical meaning of the enlarged Hilbert space may be followed
better in the three-dimensional case. The objection of Pauli may
be then removed with the help of a construction corresponding to
the approach of Lax and Phillips \cite{L&P}, which is described in
Sec. V. It allows to represent individual time-dependent solutions
of Schr\"{o}dinger equation by corresponding trajectories in
suitably extended Hilbert space. Thus, the problems concerning the
time operator in quantum mechanics may be solved completely with
the help of suitable Hilbert space modification. The possible
physical interpretation of this modification will be then
mentioned in concluding Sec. VI.

\section{Linear harmonic oscillator}

In trying to define the time operator in the case of linear
harmonic oscillator Dirac \cite{Dirac} started from the relation
fulfilled by annihilation and creation operators
\begin{equation}
  [H,a]=-\omega a, \qquad [H,a^\dagger]=\omega a^\dagger.
\end{equation}
It was possible to define operators
\begin{equation}
  {\mathcal{E}}=(a^\dagger a +1)^{-1/2} a, \qquad
   {\mathcal{E}}^\dagger = a^\dagger (a^\dagger a+1)^{-1/2},
\end{equation}
fulfilling commutation relations (\ref{[H,E]}) and corresponding
to Eq. (\ref{epx}). The operators ${\mathcal{E}}$ and
${\mathcal{E}}^\dagger$ should be, therefore, unitary. However, as
already mentioned it was shown by Susskind and Glogower \cite{SG}
that it held
\begin{equation}
   {\mathcal{E}} {\mathcal{E}}^\dagger = 1, \qquad  {\mathcal{E}}^\dagger {\mathcal{E}} = 1-|0\rangle \langle 0|;
\end{equation}
$|0\rangle \langle 0|$ being the projector onto the vacuum state.
It means that the given operators have been isometric only, but
not unitary in the standard Hilbert space; they have not
represented, therefore, a full quantum-mechanical analogue of the
classical phase exponential.

The action of ${\mathcal{E}}$ on the number-state vector basis
$\{|n\rangle\}$ reads
\begin{eqnarray}
  {\mathcal{E}} \ |n>0\rangle & \ = \ & |n-1\rangle,    \nonumber\\
  {\mathcal{E}} \ |n=0\rangle & \ = \ & 0,      \nonumber\\
  {\mathcal{E}}^\dagger \ |n\rangle & \ = \ & |n+1\rangle,
\end{eqnarray}
which can be illustrated by the following scheme:
\begin{equation}\label{retizek}
  0 \leftharpoondown |0\rangle \rightleftharpoons |1\rangle
  \rightleftharpoons |2\rangle \rightleftharpoons \dots ,
\end{equation}
where ${\mathcal{E}}$ shifts to the left and
${\mathcal{E}}^\dagger$ to the right.

The non-unitarity of the shift operators ${\mathcal{E}}$,
${\mathcal{E}}^\dagger$ has related to semi-boundedness of the
Hamiltonian spectrum. However, it has not been clear what is the
relation to the criticism of Pauli \cite{Pauli} that should be
considered more serious. It has not been clear, either, what role
has been played by the fact that until now the problem has been
discussed in the framework of one-dimensional system only.
However, before discussing the three-dimensional system, let us
mention several approaches used to define unitary phase
exponential operator for the linear harmonic oscillator.

A review of the problem, including many theoretical approaches
used so far and brief survey of experimental studies, was
published by Lynch \cite{Lynch}. Actual solution of the problem
lies in introducing a modified (extended) Hilbert space. Fain
\cite{Fain} and later independently Newton \cite{Newton} studied
an enlarged (doubled) Hilbert space consisting of two mutually
orthogonal copies of the standard space,
\begin{equation}\label{zdvojeny H}
  \mathcal H =\mathcal H_+ \oplus\mathcal H_-
\end{equation}
with the basis $\{|n,\pm\rangle\}$. Such a structure allows a
unitary phase (or time) exponential operator
${\mathcal{E}}=e^{i\Phi}$ to exist if this operator mutually links
the vacuum states of both the subspaces:
\begin{equation} \label{Phi1}
  {\mathcal{E}} = \sum_{n=0}^\infty |n,+\rangle \langle n+1,+|
    \ \  +  \ \ |0,-\rangle \langle 0,+| \ \  +  \ \ \sum_{n=0}^\infty |n+1,-\rangle \langle n,-|.
\end{equation}
Its action on the basis chain can be depicted by the scheme
\begin{equation}\label{retizek2}
  \dots \rightleftharpoons |2,-\rangle \rightleftharpoons |1,-\rangle
  \rightleftharpoons |0,-\rangle \rightleftharpoons
  |0,+\rangle \rightleftharpoons |1,+\rangle
  \rightleftharpoons |2,+\rangle \rightleftharpoons \dots ,
\end{equation}
where ${\mathcal{E}}$ shifts to the left and
${\mathcal{E}}^\dagger$ to the right.

As to the physical meaning of the doubled Hilbert space, Newton
has admitted two possible ways of Hilbert space modification, the
pseudo-spin and negative-number-state extensions \cite{Bauer} of
the standard Hilbert space, differing by the sign of Hamiltonian
on the subspace $\mathcal H_-$. However, he has considered the
subspace $\mathcal H_-$ to be merely auxiliary, the physical
meaning being ascribed to the projection onto the subspace
$\mathcal H_+$ only. On the other hand, Fain has interpreted the
doubled Hilbert space in analogy to classical mechanics; the two
subspaces have corresponded to two possible ways of connecting
phase and time, $\varphi=\varphi_0 + \omega t \ $ or $ \
\varphi=\varphi_0 - \omega t$.

Modifications of the Hilbert space proposed by other authors seem
to be more complicated. E.g., Pegg and Barnett \cite{PB} have
considered a series of finite-dimensional Hilbert spaces
${\mathcal{H}}_s$ where the phase exponential operator with the
finite basis chain $\{ |n \rangle \}_{n=0}^s$ have formed a cyclic
group; the dimension $(s+1)$ of the space has been allowed to tend
to infinity after physical results (expectation values) have been
calculated in the finite-dimensional space ${\mathcal{H}}_s$. Ban
\cite{Ban} and similarly Luis and S\'anchez-Soto
\cite{Luis+Sanchez-Soto} have tried to solve the quantum phase
problem by studying the phase-difference operator between two
systems. Another attempt to solve the given problem has been made,
e.g., by Vaccaro \cite{Vaccaro}. An extended Hilbert space
structure has been used also by Ozawa \cite{Ozawa}. A more
systematic approach has been proposed by Yu and Zhang
\cite{Yu+Zhang}. However, the proposals of Fain or Newton seem to
have been the most simple and to represent a natural way of
removing the non-unitarity of phase exponential operator.

\section{Three-dimensional harmonic oscillator}

We will derive now the exponential phase operator for
three-dimensional harmonic oscillator, which is being applied to
in many physical processes, e.g., in nuclear physics. The
Hamiltonian for such an oscillator may be written as
\begin{equation} \label{ham}
  H=\frac{({\mathbf{p}})^2}{2M}+\frac{1}{2}k \, ({\mathbf{r}})^2,
\end{equation}
where $\mathbf{r}$, $\mathbf{p}$ are position and momentum
vectors, $M$ mass of the particle, $\omega=\sqrt{\frac{k}{M}}$
angular frequency.

It is then possible to introduce vector operators
\begin{equation}
  {\mathbf{Y}}={\mathbf{p}}-iM\omega {\mathbf{r}},\qquad
       {\mathbf{Y}}^{\dagger}={\mathbf{p}}+iM\omega {\mathbf{r}},
\end{equation}
that  satisfy commutation relations
\begin{equation} \label{HY}
  [H,Y_j]=-\omega Y_j,\qquad    [H,Y_j^{\dagger} ]=\omega Y_j^{\dagger}.
\end{equation}
It holds also
 \begin{equation} \label{HY^2}
  [H,Y^2]=-2\omega Y^2,\qquad    [H,(Y^2)^\dagger ]=2\omega (Y^2)^\dagger,
\end{equation}
where
 $Y^2={\mathbf{Y}}.{\mathbf{Y}}$, $\ (Y^2)^\dagger = {\mathbf{Y}}
^\dagger . {\mathbf{Y}} ^\dagger  $.

The angular momentum $L_k = \varepsilon_{ijk} r_i p_j , \
L^2=\sum_k L_k^2 $ fulfils commutation relations
\begin{equation}
  [H,L_k]=0, \qquad [H,L^2]=0.
\end{equation}
It holds also
\begin{equation} \label{LY^2}
    [L_k, \, Y_l] \ = \ i\varepsilon_{klm} Y_m , \qquad  [L_k,Y^2]=0;
\end{equation}
operator $Y^2$ does not mix different partial waves.

It is then possible to introduce operators $Z$ and $Z^\dagger$:
\begin{eqnarray} \label{Z}
  Z &=& \frac{1}{2M}\left[(H+\omega)^2-\omega^2\left(L^2+\frac{1}{4}\right)\right]^{-1/2} Y^2 ,\nonumber\\
  Z^{\dagger} &=& (Y^2)^\dagger \frac{1}{2M}\left[(H+\omega)^2-\omega^2\left(L^2+\frac{1}{4}\right)\right]^{-1/2} ,
\end{eqnarray}
 fulfilling
\begin{equation}\label{[H,Z]}
  [H,Z]=-2\omega Z, \qquad [H,Z^\dagger]=2\omega Z^\dagger
\end{equation}
and also
\begin{equation}
  Z Z^\dagger = 1 .
\end{equation}
Operator $Z$ acts as a shift operator (similarly as
${\mathcal{E}}$ in one-dimensional case):
 \begin{eqnarray} \label{Z na bazi}
  Z |n>0,lm\rangle &=& |n-1,lm\rangle, \nonumber\\
  Z |0,lm\rangle &=& 0, \nonumber\\
  Z^\dagger |nlm\rangle &=& |n+1,lm\rangle ,
 \end{eqnarray}
where $|nlm\rangle$ are corresponding eigenfunctions of the
Hamiltonian $H$, of total angular momentum $L^2$, and of its
component $L_z$:
\begin{eqnarray} \label{baze}
  H |nlm\rangle &=& \omega (2n+l+3/2) |nlm\rangle , \nonumber\\
  L^2 |nlm\rangle &=& l(l+1) |nlm\rangle , \nonumber\\
  L_z |nlm\rangle &=& m |nlm\rangle ,
\end{eqnarray}
where $n=0,1,2,\dots; \ l=0,1,2,\dots; \ m=-l,\dots,+l.$ It means
that the operator $Z$ is an isometry operator; it violates the
unitarity when acting on vacuum state vectors $(n=0)$, similarly
as in the one-dimensional case.

\section{Doubled Hilbert space}

Similarly to the one-dimensional case, it is possible to form the
doubled Hilbert space
\begin{equation}\label{zdvojeny H -3D}
  \mathcal{H} = \mathcal{H}_+  \oplus  \mathcal{H}_-
\end{equation}
consisting of two mutually orthogonal copies isomorphic to the
standard Hilbert space. All operators act on both the subspaces in
the same way as in the standard Hilbert space.

It is convenient to use sign operator $I$ \cite{Fain,Bauer}
distinguishing the individual subspaces, which yields the
eigenvalue of $+1$ for $\mathcal{H}_+$ and $-1$ for
$\mathcal{H}_-$, and also to introduce the exchange operator $X$,
which interchanges the subspaces $\mathcal{H}_+$ and
$\mathcal{H}_-$, leaving other characteristics of the state
unchanged; it holds, e.g., $X^2=1$, $XI=-IX$, $XH=HX$,
$XL_k=L_kX$. The sign operator $I$ plays the role of a
superselection operator, as pointed out already by Bauer
\cite{Bauer} in the case of linear oscillator.

The phase exponential operator ${\mathcal{E}}=e^{i\Phi}$ may be
then defined by
\begin{eqnarray}\label{definice faze}
  {\mathcal{E}}^2  &=& \frac{1+I}{2} \ Z  \ + \ \frac{1-I}{2} \ Z^\dagger \
  + X \ (1-Z^\dagger Z) \ \frac{1+I}{2}\ , \nonumber\\
  ({\mathcal{E}}^\dagger)^2 &=& Z^\dagger \ \frac{1+I}{2} \ + \ Z \ \frac{1-I}{2} \
  + \frac{1+I}{2} \ (1-Z^\dagger Z) \ X \ ,
\end{eqnarray}
i.e., it corresponds to $Z$ and $Z^\dagger$ on the individual
subspaces and links mutually the vacuum states of both the
subspaces. The inverse formulae read
\begin{eqnarray}\label{Z pomoci faze}
  Z         &=& \frac{1+I}{2} \ {\mathcal{E}}^2  \ + \ \frac{1-I}{2} \ ({\mathcal{E}}^\dagger)^2 \ , \nonumber\\
  Z^\dagger &=& ({\mathcal{E}}^\dagger)^2 \ \frac{1+I}{2} \ + \ {\mathcal{E}}^2 \ \frac{1-I}{2}  \ .
\end{eqnarray}
The time operator may be then defined by
\begin{equation}
  T=-\Phi / \omega .
\end{equation}

Introducing sine and cosine operators
\begin{equation}\label{cos,sin}
  e^{\pm i\Phi} = \cos \Phi \pm i \sin \Phi ,
\end{equation}
one may write
\begin{eqnarray}\label{p,r pomoci faze}
  ({\mathbf{p}} - iM\omega {\mathbf{r}})^2 &=& 2M\left[(H+\omega)^2-\omega^2\Big(L^2+\frac{1}{4}\Big)\right]^{1/2} \Big(\cos 2\Phi + i I \sin 2\Phi \Big) , \nonumber\\
  ({\mathbf{p}} + iM\omega {\mathbf{r}})^2 &=& \Big(\cos 2\Phi - i I \sin 2\Phi \Big) \, 2M\left[(H+\omega)^2-\omega^2\Big(L^2+\frac{1}{4}\Big)\right]^{1/2} ,
\end{eqnarray}
i.e., the position and momentum may be expressed in terms of the
phase.

In the standard basis of the doubled Hilbert space
$\{|n,l,m;\lambda\rangle , \ \lambda=\pm 1 \},$ the sign and
exchange operators act according to
\begin{eqnarray}
  I|n,l,m;\lambda \rangle &=& \lambda |n,l,m;\lambda \rangle, \nonumber\\
  X|n,l,m;\lambda \rangle &=& |n,l,m; -\lambda \rangle,
\end{eqnarray}
and the definition of the phase exponential operator reads
\begin{eqnarray} \label{Phi3}
  e^{2i\Phi} &=& \sum_{l=0}^\infty \sum_{m=-l}^l \left( \sum_{n=0}^\infty |n,lm;+\rangle \langle n+1,lm;+| \ \  +  \ \ \right. \nonumber\\
     && \left.  +  \ \ |0,lm;-\rangle \langle 0,lm;+| \ \  +  \ \ \sum_{n=0}^\infty |n+1,lm;-\rangle \langle n,lm;-| \right), \nonumber\\
  e^{-2i\Phi} &=& \sum_{l=0}^\infty \sum_{m=-l}^l \left( \sum_{n=0}^\infty |n+1,lm;+\rangle \langle n,lm;+| \ \  +  \ \ \right. \nonumber\\
     && \left.  +  \ \ |0,lm;+\rangle \langle 0,lm;-| \ \  +  \ \ \sum_{n=0}^\infty |n,lm;-\rangle \langle n+1,lm;-| \right).
\end{eqnarray}

Let us note that the doubling procedure in the three-dimensional
case concerns the radial quantum number $n$ only, states with
different $l$ or $m$ are not mixed together. For each $l,m$ there
is the chain similar to (\ref{retizek2})
\begin{eqnarray} \label{retizekFaze3}
  \dots & \rightleftharpoons & |2,lm;-\rangle
      \rightleftharpoons |1,lm;-\rangle
      \rightleftharpoons |0,lm;-\rangle \rightleftharpoons    \nonumber\\
  & \rightleftharpoons & |0,lm;+\rangle \rightleftharpoons |1,lm;+\rangle
  \rightleftharpoons |2,lm;+\rangle \rightleftharpoons \dots ,
\end{eqnarray}
where the operator $e^{2i\Phi}$ shifts to the left and
$e^{-2i\Phi}$ to the right.

The matrix elements of the commutator $[H,e^{\pm 2i\Phi}]$ between
arbitrary states $|\chi_+\rangle, |\psi_+\rangle \in
\mathcal{H}_+$ fulfil
\begin{equation}\label{casovy operator-komutator}
  \langle \chi_+ | [ H , e^{\pm 2i\Phi} ] | \psi_+ \rangle
    = \mp 2\omega \langle \chi_+ | e^{\pm 2i\Phi} | \psi_+ \rangle.
\end{equation}
For the expectation value of $e^{\pm 2i\Phi}$ in an instantaneous
state $ |\psi_+(t)\rangle = e^{-iHt} \, |\psi_+(0)\rangle $ it
holds
\begin{equation}\label{casovy operator-ocek. hodnota}
  \langle \psi_+(t) | e^{\pm 2i\Phi} | \psi_+(t)\rangle =
    \langle \psi_+(0) | e^{\pm 2i\Phi} | \psi_+(0) \rangle \ e^{\mp 2i\omega t} .
\end{equation}
For state $|\psi_-\rangle \in \mathcal{H}_-$, these relations
differ by additional minus sign, leading to the expectation value
of $e^{\pm 2i\Phi}$ proportional to $e^{\pm 2i\omega t}$.

\section{Detailed Hilbert space structure and evolution operator}

In the doubled Hilbert space described in the preceding section,
both the subspaces have remained permanently orthogonal and
separated; evolution operator $U(t)=e^{-iHt}$ acting always in one
subspace. It has been possible to define the unitary phase
exponential operator. However, the relation to the objection of
Pauli has not been clear.

In fact, the problem of Pauli concerning a regular representation
of quantum-mechanical time evolution has been solved in the
scattering theory of Lax and Phillips \cite{L&P,L&P2} (for
summary, see \cite{R&S}). The solutions of time-dependent
Schr\"{o}dinger equation have been represented as trajectories in
extended Hilbert space consisting of the subspace
${\mathcal{D}}_-$ of incoming states and the subspace
${\mathcal{D}}_+$ of outgoing states, being connected with the
help of evolution operator $U(t)$. The following conditions have
been fulfilled
\begin{equation}\label{D+}
  U(t){\mathcal{D}}_+ \subset {\mathcal{D}}_+ \quad \forall t\geq 0,
  \qquad \bigcap_{t\geq 0} U(t){\mathcal{D}}_+ = {0}, \qquad
  \overline{\bigcup_{t\in R} U(t){\mathcal{D}}_+} = {\mathcal{H}},
\end{equation}
\begin{equation}\label{D-}
  U(t){\mathcal{D}}_- \subset {\mathcal{D}}_- \quad \forall t\leq 0,
  \qquad \bigcap_{t\leq 0} U(t){\mathcal{D}}_- = {0}, \qquad
  \overline{\bigcup_{t\in R} U(t){\mathcal{D}}_-} = {\mathcal{H}}.
\end{equation}

The same structure has been derived by Alda et al. \cite{Alda} for
the description of a purely exponential decay of an unstable
particle, the total Hilbert space being formed by the subspaces of
unstable particle, of its decay products (${\mathcal{D}}_+$), and
of colliding particles (${\mathcal{D}}_-$) producing the given
unstable particle.

Let us also mention that within the theory of Lax and Phillips,
the existence of a Hermitian (self-adjoint) time operator is a
direct consequence of the structure of the Hilbert space, i.e., of
the existence of the incoming and outgoing subspaces. Thus the
Hilbert space structure of Lax and Phillips may be considered as a
solution to the objection of Pauli. It enables to represent the
irreversible evolution already on the level of microscopic
objects. A more detailed discussion of the whole problem will be
presented elsewhere \cite{Lok-Terst}.

A similar approach may be applied also to the case of periodic
systems, e.g., to three-dimensional harmonic oscillator. The
individual Fain's subspaces ${\mathcal{H}}_+$, ${\mathcal{H}}_-$
are then formed by a series of subspaces that are mutually linked
by the action of evolution operator. The total Hilbert space
structure has the form of
\begin{eqnarray}\label{sendvicovy H}
  {\mathcal{H}} \ \ & = & \ {\mathcal{H}}_-  \oplus  {\mathcal{H}}_+  \ , \nonumber \\
  {\mathcal{H}}_\pm & = & \sum_{j=-\infty}^\infty \Big( \: {\mathcal{D}}^{(\pm)}_{-, \: j} \ + \ {\mathcal{D}}^{(\pm)}_{+, \: j} \: \Big)  \ .
\end{eqnarray}
The individual subspaces contain solutions of the time-dependent
Schr\"{o}dinger equation. E.g., the subspace
${\mathcal{D}}^{(\pm)}_{-,\: j}$ for given $j$ is formed by all
the instantaneous states belonging to the solutions of
Schr\"{o}dinger equation corresponding to one half of a period of
evolving system. Other subspaces are generated by the action of
the evolution operator according to
\begin{eqnarray}\label{H_j+1}
  U(T_0/2) \ \ {\mathcal{D}}^{(\pm)}_{-, \: j} \ \ & = & \ \ {\mathcal{D}}^{(\pm)}_{+, \: j} \ , \nonumber \\
  U(T_0/2) \ \ {\mathcal{D}}^{(\pm)}_{+, \: j} \ \ & = & \ \ {\mathcal{D}}^{(\pm)}_{-, \: j+1} \ ;
\end{eqnarray}
$T_0$ being the period of the system.

In each of the individual subspaces ${\mathcal{D}}^{(\pm)}_{\pm,\:
j}$, the phase $\Phi$ is confined to a corresponding interval:
\begin{eqnarray}
  & & {\mathcal{D}}^{(+)}_{-, \: j} \quad \ldots \quad  -2j\pi < \phi \leq \pi-2j\pi \ , \nonumber \\
  & & {\mathcal{D}}^{(+)}_{+, \: j} \quad \ldots \quad  -\pi-2j\pi < \phi \leq -2j\pi \ , \nonumber \\
  & & {\mathcal{D}}^{(-)}_{-, \: j} \quad \ldots \quad  -\pi+2j\pi < \phi \leq +2j\pi \ ,      \\
  & & {\mathcal{D}}^{(-)}_{+, \: j} \quad \ldots \quad  +2j\pi < \phi \leq \pi+2j\pi \ . \nonumber
\end{eqnarray}
The phase operator $\Phi$ exhibits monotonous behavior during time
evolution. According to (\ref{casovy operator-komutator}) -
(\ref{casovy operator-ocek. hodnota}), the expectation value of
phase $\langle\Phi\rangle$ continuously decreases with time in the
subspace $\mathcal{H}_+$ (increases in $\mathcal{H}_-$); the
expectation value $\tau$ of the time operator $T$ increases or
decreases in a similar way:
\begin{eqnarray}
  \langle\Phi\rangle (t) &=& \langle\Phi\rangle (0) \mp \omega t, \nonumber\\
  \tau (t) &=& \tau (0) \pm t.
\end{eqnarray}

\section{Conclusion}

Many physical problems have been demonstrated and solved often
with the help of linear (one-dimensional) systems. However, in
such a case actual characteristics may be simplified and some
important properties of three-dimensional systems may be hidden,
which has concerned also the problem of defining time operator.
E.g., the full sense of Fain's proposal may be seen only in the
case of the three-dimensional oscillator, as individual Hilbert
subspaces seem to correspond to different orientations of the
angular momentum vector. It might also indicate a way how to test
experimentally the physical meaning of Fain's doubling procedure.

The given scheme has been proposed to remove the problem with the
non-unitarity of phase exponential operator in the case of
harmonic oscillator. Individual subspaces have remained separated
during the whole time evolution. To solve the problem of Pauli,
both the subspaces should be formed further by a chain of other
subspaces that are mutually linked by the action of evolution
operator. The resulting structure corresponds to that proposed
earlier by Lax and Phillips for the description of scattering
phenomena. All dynamical properties of the solutions of
time-dependent Schr\"{o}dinger equation may be correctly
represented, and the phase and time operators may be defined
regularly, which enables one to distinguish between states
corresponding to different points of time evolution.

Therefore, the use of the more complex Hilbert space structure
seems to be not only formal, as sometimes stated, but also
physically reasonable and practically necessary. The use of the
structure described in the preceding enables us to represent the
time evolution of periodic and non-periodic systems on the similar
basis.


\end{document}